\documentclass[12pt]{iopart}

\usepackage{graphicx} 

\begin{document}

\title{Elastic precursor of the transformation from glycolipid-nanotube to -vesicle}

\author{
T Fujima$^{1}$ \footnote{Present address: EcoTopia Science Institute, Nagoya University, Furo-cho, Chikusa-ku, Nagoya 464-8603, Japan. E-mail: fujima{\_}t@plasma.numse.nagoya-u.ac.jp}, 
H Frusawa$^2$\footnote{Present address: Soft Matter Laboratory, Kochi University of Technology, Tosa-Yamada, Kochi 782-8502, Japan. E-mail: frusawa.hiroshi@kochi-tech.ac.jp}, 
H Minamikawa$^3$, K Ito$^1$ and T Shimizu$^3$
}

\address{$^1$ Graduate School of Frontier Sciences, University of Tokyo, 5-1-5 Kashiwa-no-Ha, Kashiwa 277-8561, Japan}
\address{$^2$ Department of Applied Physics, University of Tokyo, 7-3-1 Hongo, Bunkyo-ku, Tokyo 113-8656, Japan}
\address{$^3$ Nanoarchitectonics Research Center (NARC), National Institute of Advanced Industrial Science and Technology (AIST), Tsukuba Central 5, 1-1-1 Higashi, Tsukuba 305-8565, Japan}

\date{\today}

\begin{abstract}
By the combination of optical tweezer manipulation and digital video microscopy, the flexural rigidity of single glycolipid "nano" tubes has been measured below the transition temperature at which the lipid tubules are transformed into vesicles.
Consequently, we have found a clear reduction of the rigidity obviously before the transition as temperature increasing.
Further experiments of infrared spectroscopy (FT-IR) and differential scanning calorimetry (DSC) have suggested a microscopic change of the tube walls, synchronizing with the precursory softening of the nanotubes.
\end{abstract}

\pacs{62.25.+g, 81.16.Fg, 87.16.Ka, 81.07.Nb, 62.20.Dc, 87.16.Dg}

\maketitle

\section{Introduction}

Among the most promising supramolecular structures is the lipid tubule, a spontaneous aggregate of chiral and amphiphilic molecules~\cite{yager1,yager2,kunitake}.
The structure is a hollow cylinder of lipid bilayer, and the typical diameter is of sub-$\mu$m to $\mu$m scale.
Recently, it has also been found that a variety of glycolipids forms fine tubules with diameters of tens of nanometers~\cite{cardanol1}.
These lipid tubules, unlike carbon nanotubes, have a unique characteristic that both sides of the lipid-bilayer walls, internal and external, are hydrophilic.
Therefore, the quasi-one-dimensional nanopore of the glycolipid tubule has potentialities of chemical reactions, the transferal of biomolecules and so on.
Partly motivated by these fascinating possibilities, the spontaneous formation of lipid tubules has been studied both experimentally~\cite{oda,yang,bassereau,thomas,fuhrhop,nolte,menger,meijer} and theoretically~\cite{prost,lipid_tube_th,lipid_tube_th2}, and it is now possible to tune their morphology~\cite{tube_tune,cardanol2,tube_tune2}. 

Cellular microtubules, on the other hand, have similarities to the lipid nanotube in that the diameter is of the same order, and that the inner space is hydrophilic since they are made of proteins.
The microtubules play roles of cytoskeleton or flagellum, suggesting another application of both microtubules and the lipid nanotubes: they are regarded as elastic nano-rods, and are good candidates for the components of micro- or nano-structured materials.
Accordingly, it is significant to study fabrication mechanics, i.e. mechanical property of the single nanotubes, as well as to investigate the self--assembly mechanism.

The stiffness of microtubules, cytoskeleton, has been measured by various techniques~\cite{microtubule_mecha1,microtubule_mecha2,microtubule_mecha3,microtubule_mecha4,microtubule_ana1,microtubule_ana2,microtubule_nanomecha}; for example, a latest result using AFM has shown that the rigidity of microtubules decreases by increasing temperature~\cite{microtubule_nanomecha}, correspondingly to a certain change of the protein wall.
By contrast, however, there have been few experiments for the elasticity of the lipid nanotubes except the preliminary result by some of the present authors~\cite{angew}.
We then performed the rigidity measurements with the use of optical tweezers, precisely varying the temperature below the tube--vesicle transformation.
Further experiments, Fourier-transformed infrared spectroscopy (FT-IR) and differential scanning calorimetry (DSC), were also carried out from a perspective of microscopic structure.
Here we will report the new finding that the single glycolipid-nanotubes exhibit a precursory reduction of their rigidity before transforming to vesicles.

\section{Material}

The lipid tubules, used here, consist of a renewable-resource-based synthetic glycolipid (cardanyl-$\beta$-D-glucopyranoside).
We prepared the tubules through self-assembly of the lipids under a mild condition as follows:
The powdery sample was dissolved in water by vortexing at 100 $^\circ$\hspace{-.13em}C for 30 min (ca. $3\times 10^{-4}$ wt\%).
Then one-week of the incubation at room temperature was sufficient to have the supramolecular assemblies which kept the nanotube structure for years.
It is found from Proton NMR analysis (JEOL LA 600 NMR spectrometer, 600 MHz, CDCl3) of the tubes that the lipid contains different unsaturations of the alkyl chains: saturated one (7\%), monoene (80\%), diene (12\%) and triene (1\%).
We have found from the transmission electron microscopy (TEM) images that the outer and inner diameters of the tubules are 50 and 10 nm with narrow dispersion, respectively, which is in fact a nanotube~\cite{angew}.

\begin{figure}[hbtp]
\begin{center}
\includegraphics{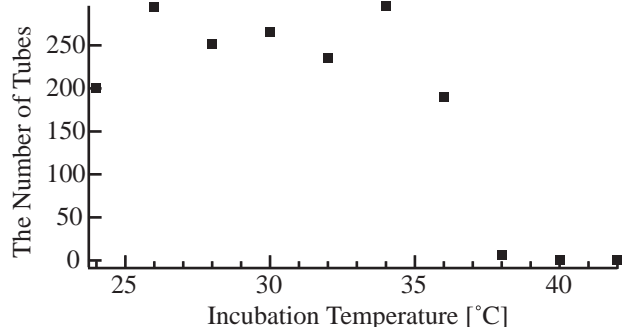}
\caption{Incubation temperature versus the number of tubes which were counted and averaged in a specified area after one-week incubations, using the differential interference microscopy. The grey line is to guide the eye.}
\end{center}
\end{figure}

The structural transition temperature was determined by the incubation temperature dependence of the tube forming ability and the tube to vesicle transition.
We counted, by the differential interference microscopy, the number of nanotubes in a fixed volume after one-week incubations keeping the temperatures between 24 $^\circ$\hspace{-.13em}C and 42 $^\circ$\hspace{-.13em}C as shown in figure 1.
This indicates that the transformation temperature is located between 36 $^\circ$\hspace{-.13em}C and 38 $^\circ$\hspace{-.13em}C, that is consistent with the observation of the transformation from the nanotubes to vesicles when raising the temperature of the nanotube dispersion.

\section{Methods}

\subsection{Rigidity Measurement}

To measure the flexural rigidity of the single lipid-nanotubes, we used the optical tweezers system (Sigma Koki LMS-46755).
The process was three fold.

First, we placed a drop of the aqueous dispersion of the lipid nanotube on a glass slide, then enclosed it with coverslip and resin.
The concentration used was the same as that in the sample preparation.
We have confirmed that nanotubes are completely isolated and form no bundles in the density~\cite{angew}.
The sample was somewhat pressed between the coverslip and the slide.
As a result, some nanotubes adhered firmly to the glass surface especially at their ends, and it became easy to find nanotubes fixed at only one end.
Next, the one-end-free nanotube was bent similarly to microtubules~\cite{microtubule_ana1,microtubule_ana2} as follows:
Using optical tweezers controllable with the spacial resolution of 0.1 $\mu$m,
we nudged the free end in the direction perpendicular to the long axis of the tube so that the nanotube could be bent to an adequate angle, a few tens of degrees.
Finally, the laser beam was switched off after the bending process, and correspondingly the bow-shaped nanotube started relaxing to its initial straight form (see figure 2).
The relaxation process was recorded as a movie at a rate of 30 frames per second.

\begin{figure}[htbp]
\begin{center}
\includegraphics[scale=0.75]{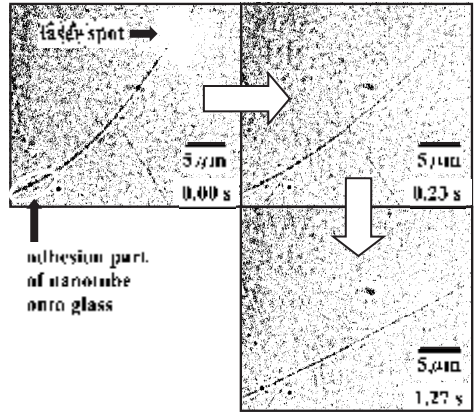}
\caption{Pictures from bow-shape to the initial straight form of a single nanotube. The temperature was fixed at 27 $^\circ$\hspace{-.13em}C.}
\end{center}
\end{figure}

In figure 3, a typical result of the shape relaxation is shown as the time dependence of $y(l, t)$, where $y$ is ordinate perpendicular to the initial straight line, and $l$ the contour length from the adhesion point to the trapped point; $l=23 \mu$m in the present case. The solid line in figure 3 represents the curve fitted by the following single exponential:
\begin{eqnarray}
&&y(l,t)=y(l,0) \exp{ \left( -t/ \tau \right) } \\
&&\tau=\frac{\pi \eta L^4}{60K\ln{(L/2d)}} \left\{ \left( \frac{l}{L} \right)^5-10 \left( \frac{l}{L} \right)^3+20 \left( \frac{l}{L} \right)^2 \right\},\nonumber\\
\end{eqnarray}
where $\tau$ is the relaxation time, $\eta$ the viscosity of surrounding water, $K$ the flexural rigidity, and $L$ the contour length  between the fixed point and the free end, and $d$ the outer diameter of the present nanotube.
The above equations come from the balance condition between elastic and hydrodynamic force~\cite{microtubule_ana2}.

\begin{figure}[hbtp]
\begin{center}
\includegraphics[scale=0.6]{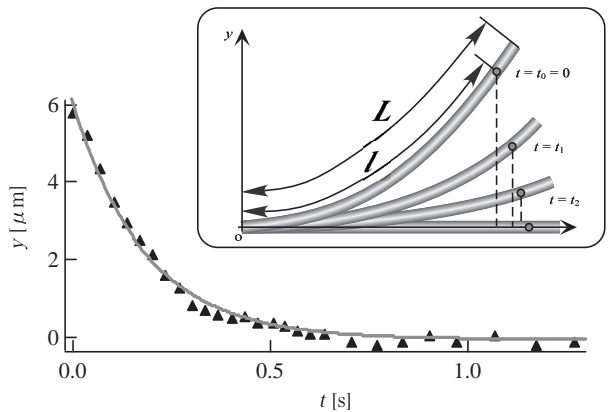}
\caption{The relaxation data analyzed on figure 2. We plot by the triangles time series of ordinates defined in the schematic of the inset, and the solid line represents the best fitting result by equations (1) and (2). In this case, evaluating $\tau$ and $K$ gives 0.23 [s] and 1.3 [$10^{-22}$ Nm$^2$] when $l$=22.6 $\mu$m, $L=26\,\mu\mathrm{m}$ and $d=50$ nm.}
\end{center}
\end{figure}

We take the parameter $d$ equal to 50 nm, the typical value of the nanotube~\cite{cardanol1}, in the rigidity analyses, since the variation of tube diameter is irrelevant to the rigidity evaluation based on the equations (1) and (2); even doubling or reduction by half of the diameter changes the stiffness by about 10\%.
The contour length $L$, on the other hand, was evaluated within the accuracy of 0.5$\mu m$ for each temperature, because the time series of pictures in the movie of structural relaxation determined definitely the locations of both the fixed point, working as a fulcrum, and the other free ened. It was also confirmed that the contour length was invariant with change in temperature.

The rigidity $K$ was obtained from the fitting result in the temperature range of 20 -- 30 $^\circ$\hspace{-.13em}C.
We raised the temperature of the aqueous dispersion of the nanotubes from the lower end to the other, keeping the rate 0.1 $^\circ$\hspace{-.13em}C min$^{-1}$ within the accuracy of 0.1 $^\circ$\hspace{-.13em}C throughout.
The measuring time at each temperature was so short that the temperature change during every movie acquisition was negligible.

\subsection{FT-IR}

We investigated the temperature dependence of FT-IR spectra for the aqueous nanotube dispersion, using the spectrometer (SHIMAZU FTIR-8700) with the ATR-type cell (DuraSampl IR II).
Temperature was controlled within 0.1 $^\circ$\hspace{-.13em}C and varied in the same range as that of the elasticity measurements.
The measurements were isothermally performed at every 1.0 degree in the temperature range, and the heating rate was 0.1 $^\circ$\hspace{-.13em}C min$^{-1}$ equally to that in the rigidity measurements.
To get a sufficient signal intensity and the wavenumber resolution, we raised the sample concentration by osmotic compression to 10 times as high as that used in the rigidity measurements and obtained the resolution of 3.85 cm$^{-1}$.
Even after the condensation, the optical microscopy has confirmed that the nanotubes are still present.

\subsection{DSC}

We also performed thermoanalysis by the differential scanning calorimeter (Seiko Instruments DSC6100).
The temperature was changed in the range of 20 -- 40 $^\circ$\hspace{-.13em}C at a heating rate of 1.0 $^\circ$\hspace{-.13em}C/min.
In order to obtain the sufficient signal intensity of DSC, the concentration of the lipid solution was required to be more than 5 wt\% which is too high for nanotubes to keep their supramolecular structure.
This measurement, therefore, was performed for simple solution of the lipid and does not reflect the supramolecular structure of the lipid nanotubes but the nature of the lipid molecules themselves.

\section{Results and Discussions}

\subsection{Rigidity}

Figure 4 shows the temperature dependence of the flexural rigidity, indicating that the stiffness drops off in the range 23 -- 27 $^\circ$\hspace{-.13em}C with increasing the temperature and the ratio of the minimum to maximum is about half, though we have observed no sign of the macroscopic transformation.
The results for different nanotubes were reproducible within the error bars.
This supports the assumption that the friction due to the glass slides, which should vary considerably as a result of the tube arrangement on the prepared slide, is much less tha tha viscosity of water.
It should also be noted that the above variations of both the tube-diameter $d$ and tube-length $L$ leads to deviations within the error bars.

\begin{figure}[htbp]
\begin{center}
\includegraphics[scale=0.5]{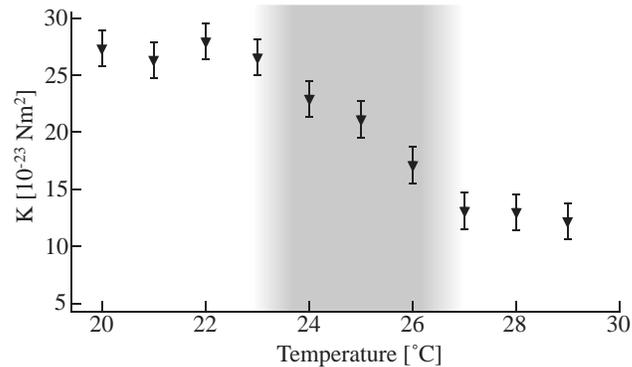}
\caption{Temperature dependence of the flexural rigidity $K$. The triangles represent the average, and the error bars were estimated from measurement-variations and resolution of the movies. Filled zone marks the temperature range where stiffness is decreasing.}
\end{center}
\end{figure}

The FT-IR and DSC measurements provided meaningful results for the investigation of whether this precursory softening before the tube-vesicle transformation is a generic phenomenon reflecting a change of microscopic structure.

\subsection{FT-IR}

Figure 5 represents the IR spectra when raising the temperature. As clearly shown in the figure, the absorption band around 1460 cm$^{-1}$ changes its shape from twin peaks to single one between the spectra of 26 and 27 $^\circ$\hspace{-.13em}C where the elasticity reduction also occurs as mentioned above.
Up to now, it has been found that this absorption band arises from the scissoring motion of CH${_2}$ unit in hydrocarbon chains and is sensitive to the molecular in-plane packing of lipid bilayers~\cite{FTIR_lateral}.
The twin- to single- peak change has also been reported for the lamellar structure of other lipids when temperature increased~\cite{FTIR_pack}; in this case, while the twin-peak spectra is due to the tight packing of lipids in the crystal phase, the single-peak is associated with the gel phase.
The previous reports suggest that the present result is ascribable to a microscopic change of the tube walls inducing the nanotube softening.

\begin{figure}[htbp]
\begin{center}
\includegraphics[scale=0.66]{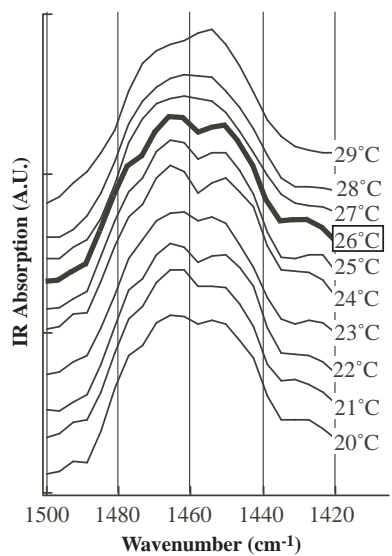}
\caption{Temperature dependence of FT-IR spectra.
For visual clarity, each spectrum is vertically shifted at even intervals of the temperatures.
The spectrum at 26 $^\circ$\hspace{-.13em}C is emphasized by thicker line, because the peak shape changes around here.}
\end{center}
\end{figure}


\subsection{DSC}

It is found from Fig. 6 that there is an endothermic peak around 25.5 $^\circ$\hspace{-.13em}C where the tube stiffness has started to decrease as shown in figure 4.
This result indicates that the nanotube softening and the molecular packing change in the tube walls are based on a thermodynamic phenomenon.
Also, the DSC peak is broad similarly to the gradual softening in figure 4.
Although the present DSC result is obtained from condensed solutions of the glycolipids without forming nanotubes as described in section 3-3, the blunt thermo--anomaly below the tubule-vesicle transformation temperature has also been reported in a similar way for other lipid tubules with large diameters~\cite{kumar}.

\begin{figure}[htbp]
\begin{center}
\includegraphics[scale=0.75]{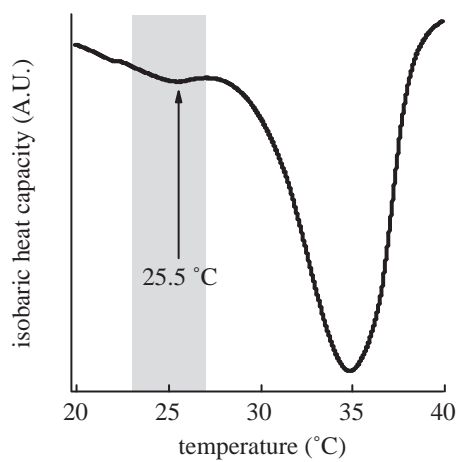}
\caption{DSC result for the lipid solution, which indicates some thermodynamic change in the single lipid molecules around 25.5 $^\circ$\hspace{-.13em}C. Filled zone marks the temperature range where the stiffness is decreasing (see figure 4).}
\end{center}
\end{figure}


\section{Concluding Remarks}

We have thus unveiled the key temperature located around 26 $^\circ$\hspace{-.13em}C where the reduction in stiffness (see figure 4), the change in the IR-spectrum in the 1460 cm$^{-1}$ band (see figure 5) and endothermic peak (see figure 6) have been observed synchronously for the glycolipid nanotubes.
Combining all the results suggests that the softening of the lipid nanotubes is linked with the microscopic change of constituent lipids.
It is then reasonable to say that the gentle decrease in the elasticity seen in figure 4 is due to the broad endothermic peak in figure 5.
The present unsharpness may be explained by considering that the sample is a mixture of different types of alkyl chains as mentioned above.
That is, although each pure lipid has fast equilibration time, the changes are smeared and appear to be gradual due to the averaging of mixed lipids.

To summarize, we investigated the temperature dependence of the nanotube made of the cardanol-glycolipid below the tube-vesicle transformation temperature.
Elasticity measurements have revealed a precursory phenomenon of supramolecular structure transformation in the lipid nanotube system.
That is, the decrease of the stiffness with increasing temperature while lipids maintain the tubule structure.
The results of FT-IR and DSC further suggest that the softening is associated with a microscopic change in the lipid wall which we are unable to detect using optical microscopy.

\ack
We thank S. Kubo, T. Masui, A. Fukagawa, T. Ikehara and T. Nishi for valuable discussions and helps. This research was supported by the JST-CREST. One of the authors (T. F.) acknowledges also the partial support from the JSPS Research Foundation for Young Scientists.


\end{document}